# Consciousness:
# Here, There but Not Everywhere

by

Giulio Tononi[1] and Christof Koch[2]

[1]Department of Psychiatry, University of Wisconsin, Madison WI USA

[2]Allen Institute for Brain Science, Seattle, WA USA

May 26, 2014

Abstract

The science of consciousness has made great strides by focusing on the behavioral and neuronal correlates of experience. However, correlates are not enough if we are to understand even basic facts, for example, why the cerebral cortex gives rise to consciousness but the cerebellum does not, though it has even more neurons and appears to be just as complicated. Moreover, correlates are of little help in many instances where we would like to know if consciousness is present: patients with a few remaining islands of functioning cortex, pre-term infants, non-mammalian species, and machines that are rapidly outperforming people at driving, recognizing faces and objects, and answering difficult questions. To address these issues, we need not only more data, but also a theory of consciousness – one that says what experience is and what type of physical systems can have it. Integrated Information Theory (IIT) does so by starting from conscious experience itself via five phenomenological axioms of existence, composition, information, integration, and exclusion. From these it derives five postulates about the properties required of physical mechanisms to support consciousness. The theory provides a principled account of both the quantity and the quality of an individual experience (a quale), and a calculus to evaluate whether or not a particular system of mechanisms is conscious and of what. Moreover, IIT can explain a range of clinical and laboratory findings, makes a number of testable predictions, and extrapolates to a number of unusual conditions. The theory vindicates some intuitions often associated with panpsychism - that consciousness is an intrinsic, fundamental property, is graded, is common among biological organisms, and even some very simple systems may have some of it. However, unlike panpsychism, IIT implies that not everything is conscious, for example aggregates such as heaps of sand, a group of individuals or feed-forward networks. Also, in sharp contrast with widespread functionalist beliefs, IIT implies that digital computers, even if their behavior were to be functionally equivalent to ours, and even if they were to run faithful simulations of the human brain, would experience next to nothing.

## Consciousness:

### Here, There and Everywhere?

I know I am conscious: I am seeing, hearing, feeling something *here*, inside my own head. But is consciousness – subjective experience - also *there,* not only in other people's heads, but also in the head of animals? And perhaps *everywhere,* pervading the cosmos, as in old panpsychist traditions and in the Beatles' song? While these kinds of questions may seem scientifically inappropriate, we argue below that they can be approached in a principled and testable manner. Moreover, obtaining an answer is urgent, not only because of difficult clinical cases and in our interactions with other species, but also because of the advent of machines that are getting closer to passing the Turing test – computers programmed to perform many tasks as well as us, and often far better than some brain-damaged patients.

### Here

That I am conscious, here and now, is the one fact I am absolutely certain of – all the rest is conjecture. This is, of course, the gist of the most famous deduction in Western thought, Descartes' *je pense, donc je suis*. Everything else – what I think I know about my body, about other people, dogs, trees, mountains, and stars, is inferential. It is a reasonable inference, corroborated first by the beliefs of my fellow humans and then by the intersubjective methods of science. Yet consciousness itself - the central fact of existence – still demands a rational explanation.

The past two centuries of clinical and laboratory studies have revealed an intimate relationship between the conscious mind and the brain, but the exact nature of this relationship remains elusive. Why does the brain generate consciousness but not the liver or the heart, as previous cultures believed? Why certain parts of the brain and not others? Why is consciousness lost in some stages of sleep? Why does red feel like red and not like the sound of a violin? Is consciousness just an epiphenomenon, or does it have a function? Can computers be conscious? Could a system behave like us and yet be devoid of consciousness – a zombie? Such questions seem to resist the empirical, reductionist approach that has been so successful for other aspects of the natural world. Nevertheless, thanks to experimental and theoretical progress in the past decades (Koch 2004, Baars and Gage 2010, Dehaene and Changeux 2011, Boly, Baars et al. 2013) (Laureys, Tononi et al. 2009), we are finally in a better position to understand which systems under which conditions can give rise to consciousness. That is, the study of consciousness is becoming a science. In doing so, it is leaving behind the defeatist dictum of the physiologist Emil du Bois-Reymond, *ignoramus et ignorabimus* (we don't know and never will), espousing instead the upbeat maxim of the mathematician David Hilbert, *Wir müssen wissen – wir werden wissen* (we must know and we will).

### There

We usually grant consciousness to others - of the same kind we experience in the privacy of our own mind – if they can tell us what they feel, or if they look and behave more or less like us. However, we become less and less confident in attributing consciousness to people who cannot talk about their experiences, such as infants and young children, or severely brain injured patients. Many assume that animals closely related to *homo sapiens* – apes and other primates – are conscious, though presumably less than we are, based on the similarity of their behavior and their brain. But should we attribute experience to all mammals,[i] to all vertebrates, to invertebrates such as cephalopods and bees, or even to all multi-cellular animals? What about cultured organoids that mimic the cellular organization of the developing human brain (Lancaster, Renner et al. 2013)? And finally, what about the sophisticated machines that run software designed to substitute for conscious humans in many complicated tasks?

### Behavioral correlates of consciousness (BCC) and reportability

Traditionally, we assess consciousness by observing behavior (Fig. 1A). If someone is awake and acts meaningfully, we have little doubt he is conscious. If he speaks, and especially if he can answer questions about what he is conscious of, we are fully confident. In the laboratory, the ability to report one's experiences has become the gold standard for judging the presence of consciousness. Reportability is often reduced to a binary forced choice, in which the subject only pushes one of two buttons for "seen" vs. "not seen," or "angry" vs. "happy face". One can also ask how confident one is that one has seen something. These kinds of meta-cognitive reports can also be obtained from trained monkeys or other animals, with so many similarities to our own reports that we have little doubt as to the presence of consciousness (Cowey and Stoerig 1995).

But behavior can be misleading: a person may walk and speak in her sleep, yet it is quite dubious whether she is experiencing anything. Or a person can be asleep, immobile, silent, and unresponsive, yet she may be dreaming - vividly conscious of an imaginary environment. In such cases, reportability can be used as retrospective evidence of consciousness, by waking up the sleeper to obtain a "dream report." However, reportability, too, can be problematic. Since we obviously experience things in dreams whether or not we are woken up to report them, we should accept the possibility that in certain situations consciousness can be present even if it cannot be reported (Lamme 2006, Block 2007). Moreover, insisting on reportability elevates language to a king-maker role, which makes inferring consciousness in non-verbal infants, preterm babies, fetuses, or animals problematic.[ii] Clearly, if we want to understand what is really going on, we must investigate the brain mechanisms that underlie consciousness.

### Neural correlates of consciousness (NCC)

The NCC have been defined as the minimal neural mechanisms that are jointly sufficient for any one conscious percept (Fig. 1B; (Koch and Crick 1990, Crick and Koch 1998, Koch and Crick 2000). Every experience will have associated NCC: one for seeing a red patch, another one for hearing a high C. Inducing the NCC by manipulating the relevant neuronal populations via magnetic stimulation, optogenetics or other means will give rise to the associated conscious percept, and interfering with the NCC by disabling the underlying neural circuits will eliminate the percept.

The NCC are assessed at first by determining which aspects of neural function change depending on whether a subject is conscious or not, as established using behavioral reports. This can be done by considering a global change in the state of consciousness, as when awareness is lost during deep sleep or general anesthesia (Tononi and Laureys 2009). Or it can be done by considering changes in a particular content of consciousness, as when a subject's awareness of a particular stimulus is experimentally manipulated ("seen" vs. "not seen"). In optimally controlled experiments, the stimulus and the behavioral report (such as a button press) are kept constant while the subject sometimes sees the percept and sometimes does not (Logothetis 1998, Dehaene and Changeux 2011, Mudrik and Koch 2013). Once a particular NCC has been sufficiently validated, it can be used to extrapolate to situations where reports are not available. Both functional brain imaging in magnetic scanners as well as large-scale EEG recordings from outside the skull have been put to use to track down the footprints of consciousness in the brain of healthy adult observers. Popular candidates include strong activation of higher-order fronto-parietal cortices (Fig. 1B), high-frequency electrical activity in the gamma range (35-80 Hz), and the occurrence of an EEG event known as the P300 wave (Tononi and Laureys 2009, Dehaene and Changeux 2011). However, there is still no consensus on whether any of these signs can be treated as reliable "signatures" of consciousness. In particular, there can be consciousness without frontal cortex involvement (Mataró, Jurado et al. 2001, Goldberg, Harel et al. 2006, Frässle, Sommer et al. 2014), gamma activity without consciousness(Engel and Singer 2001), such as during anesthesia (Imas, Ropella et al. 2005, Murphy, Bruno et al. 2011), and consciousness without a frontal P300, for example during dreaming sleep (Cote, Etienne et al. 2001, Takahara, Nittono et al. 2002). Moreover, it is likely that many of the signatures proposed as possible NCC may actually be correlates of neural activity that is needed leading up to a conscious percept (Aru, Axmacher et al. 2012, Pitts, Martínez et al. 2012), or for giving a report following a conscious percept (Goldberg, Harel et al. 2006, Pitts, Martínez et al. 2012, Frässle, Sommer et al. 2014), rather than for having an experience. Finally, NCC obtained in healthy adults may or may not apply to brain damaged patients, to infants, to animals very different from us, and certainly not to machines (Fig. 2).

### Patients and infants

Patients with widespread cortical or thalamic damage pose a poignant challenge. Emergency room personnel quickly evaluate the severity of a head injury behaviorally by assigning a number to a patient's auditory, visual, verbal and motor functions as well as communication and arousal level. Various NCC, such as the presence of a P300 wave in response to a non-standard stimulus, are increasingly being used to complement the behavioral assessment and occasionally modify the diagnosis. In some cases, NCC can be decisive. Thus, if a patient who lies mute and immobile can, however, respond to commands by appropriately activating certain brain areas, it is fair to conclude that she is conscious (Owen, Coleman et al. 2006). Yet even the NCC may be ambiguous. For example, the P300 wave is absent in many minimally conscious patients and even in some brain-damaged patients who can communicate (King, Sitt et al. 2013). And what should one make of patients in whom, amidst widespread destruction and inactivity, one or a few isolated cortical areas may show signs of metabolic activation and of electrophysiological "markers" of consciousness (Schiff, Ribary et al. 2002)? Is an island of functioning brain tissue sufficient for generating a limited kind of awareness, maybe just awareness of sound or of pain? In other words, "what is it like" to be a brain island, if it feels like anything at all? And how big must the island be to qualify?

By the same token, what is it like to be a newborn baby with an immature brain and restricted connectivity among cortical structures (Lagercrantz and Changeux 2009)? Again, considering NCC can be helpful: for example, a wave resembling the P300 wave has been reported in 6 to 16 months old infants, although weaker, more variable and delayed than in adults (Kouider, Stahlhut et al. 2013). But does this mean that newborn and pre-term babies or even fetuses experience nothing because they do not show a P300?

### Animals

The problem becomes even more acute when turning to other species. The study of consciousness in nature has been hindered for centuries by a strong belief in human exceptionalism. Yet the range and complexity of animal behavior has laid rest to this belief, at least among biologists. This is particularly true for mammals. In psychophysical tasks involving simple button presses, trained macaque monkeys act very similar to human volunteers, including signaling when they don't see anything (Cowey and Stoerig 1995). Visual recognition of self, meta-cognition (knowing one's mind), theory of mind, empathy and long-range planning have all been demonstrated in primates, rodents and other orders (Smith, Couchman et al.).

It is also difficult to find anything exceptional about the human brain (Hawrylycz, Lein et al. 2012). Its constitutive genes, synapses, neurons and other cells are similar to those found in many other species. Even its size is not so special, as elephants, dolphins and whales have even bigger brains (Herculano-Houzel 2012). Only an expert neuroanatomist, armed with a microscope, can tell a grain-sized piece of neocortex of a mouse from that of a monkey or a human. Biologists emphasize this structural and behavioral continuity by distinguishing between *non-human* and *human* animals (Huxley 1872). Given this continuity, it seems unjustified to claim that only one species has consciousness while everybody else is devoid of experience, is a zombie. It is far more likely that all mammals have at least some conscious experiences, can hear the sounds and see the sights of life.

As we consider species that are progressively further removed from *homo sapiens* in evolutionary and neuronal terms, the case for consciousness becomes more difficult to make. Two observations, one relating to complexity of behavior and another one to complexity of the underlying nervous system, are critical. First, ravens, crows, magpies, parrots and other birds, tuna, coelacanths and other fish, octopuses and other cephalopods, bees and other members of the vast class of insects are all capable of sophisticated, learnt, non-stereotyped behaviors that we associate with consciousness if carried out by people (Dawkins 1998, Griffin 2001, Edelman and Seth 2009). Darwin himself set out "to learn how far the worms acted consciously" and concluded that there was no absolute threshold between "lower" and "higher" animals, including humans, which would assign higher mental powers to one but not to the other (Darwin 1887). Second, the nervous systems of any of these creatures display a vast and ill-understood complexity. The bee contains about 800,000 nerve cells whose morphological and electrical heterogeneity rivals that of any neocortical neuron. These cells are assembled in highly nonlinear feedback circuits whose density is up to ten times higher than that of neocortex (Koch and Laurent 1999). Thus, it is unlikely that neural signatures of consciousness that have some validity in humans and other mammals will apply to invertebrates.

On the other hand, the lessons learnt from studying the behavioral and neuronal correlates of consciousness in people must make us cautious about inferring its presence in creatures very different from us, no matter how sophisticated their behavior and how complicated their brain. Humans can perform seemingly sophisticated behaviors - recognizing whether a scene is congruous or incongruous, controlling the size, orientation, and strength of how one's finger should grip an object, doing simple arithmetic, detecting the meaning of words, or rapid keyboard typing – in a seemingly non-conscious manner (Hassin, Uleman et al. 2005, Kouider and Dehaene 2007, Berlin 2011, Mudrik, Breska et al. 2011, Sklar, Levy et al. 2012, Hassin 2013). When a bee navigates a maze, does it do so like when we consciously deliberate whether to turn right or left, or rather like when we type on a keyboard? Similarly, consider that an extraordinarily complicated neuronal structure in our brain, the cerebellum, home to 69 of the 86 billion nerve cells that make up the human brain (Herculano-Houzel 2012), apparently has little to do with consciousness. Patients that lose part or nearly all of their cerebellum due to stroke or other trauma show ataxia, slurred speech, and unsteady gait (Lemon and Edgley 2010) but do not complain of a no loss or diminution of consciousness. Is the bee's brain central complex more like the cerebellum or more like the cerebral cortex with respect to experience? Thus, the extent to which non-mammalian species share with us the gift of subjective experience remains hard to fathom.[iii]

### Machines

Difficulties in attributing sentience become even more apparent when considering digital computers. These have a radically different architecture and provenance than biological creatures shaped by natural selection. Due to the relentless decrease in transistor size over the past 50 years and the concomitant exponential increase in computational power and memory capacity, present-day computers executing appropriate algorithms outperform us in many tasks that were thought to be the sole prerogative of the human mind. Prominent examples include IBM's Deep Blue that beat the reigning chess world master in 1997; another IBM computer, Watson, capable of answering questions posed in spoken English that won the quiz show *Jeopardy* in 2011; smart phones that answer questions by speech; Google's driverless cars that have logged more than half a million km on open roads; and machine vision algorithms for face detection in security and commercial applications (Kurzweil 2012). People playing chess, supplying a meaningful answer to a given question, driving a car or picking out a face are assumed to be conscious. But should we say the same for these digital creatures?

**Integrated Information Theory (IIT)**

Clearly, as we move away from people, BCC and NCC become progressively less helpful to establish the presence of consciousness. Also, even in the normal human brain, we need to understand *why* and *how* certain structures give rise to experience (the cerebral cortex), while others do not (the cerebellum), and why they do so under certain conditions (wake, dreams) and not others (deep sleep, seizures). Some philosophers have claimed that the problem of explaining how matter can give rise to consciousness may forever elude us, dubbing it the *Hard* Problem (Chalmers 1996, Block, Flanagan et al. 1997, Shear 1999). Indeed, as long as one starts from the brain and asks how it could possibly give rise to experience – in effect trying to "distill" mind out of matter (Mcginn 2000), the problem may indeed remain not only hard, but outright impossible to solve. But things may be less hard if one takes the opposite approach: start from consciousness itself, by identifying its essential properties, and then ask what kinds of physical mechanisms could possibly account for them. This is the approach taken by Integrated Information Theory (IIT) (Tononi 2004, Tononi 2008, Tononi 2012), an evolving framework that provides a principled account for what it takes for consciousness to arise, offers a parsimonious explanation for the empirical evidence, makes testable predictions, and permits inferences and extrapolations.[iv]

**Axioms: Essential phenomenological properties of consciousness**

Taking consciousness as primary, IIT first identifies *axioms* of experience (Fig. 3, left), then derives a set of corresponding *postulates* (Fig. 3, right) about the nature of the underlying physical mechanisms (Tononi 2012, Oizumi, Albantakis et al. 2014). The axioms of IIT are truths about our own experience that are assumed to be self-evident. They include existence, composition, information, integration, and exclusion.

*Existence*. Consciousness *exists*: my experience just *is* – indeed, that my experience here and now exists, is the only fact I can be absolutely sure of. Moreover, my experience exists from its own *intrinsic* perspective, independent of external observers.

*Composition*. Consciousness is *structured*: each experience is composed of many *phenomenological distinctions*. Within the same experience, I can see, for example, left and right, red and blue, a triangle and a square, a red triangle on the left, and so on.

*Information*. Consciousness is *differentiated*: at any moment in time, each experience is *the specific way it is* (a *specific* set of phenomenological distinctions), differing in its *specific* way from other possible experiences. Thus, an experience of pure darkness and silence is what it is because, among other things, it is not filled with light and sound, color and shapes, and so on. Consider all the frames of all possible movies: these are but a small subset of all possible experiences.

*Integration*. Consciousness is *unified*: each experience is *irreducible* to non-interdependent components. Thus, I experience a whole visual scene, not the left side of the visual field independent of the right side (and vice versa). For example, the experience of seeing the word "HONEYMOON" written in the middle of a blank page is irreducible to an experience of seeing "HONEY" on the left plus the experience of seeing "MOON" on the right. What I see is the whole "HONEYMOON." Similarly, seeing a red triangle is irreducible to seeing a grey triangle plus the disembodied color red.

*Exclusion*. Consciousness is *singular*, in content, and spatio-temporal grain: there is *no superposition* of multiple experiences, with less or more content, flowing at once at faster or slower speeds. Thus, in addition to the experience I have right now, say, of being in my room and seeing my legs on the bed, there are no further superimposed experiences with either less content – say, one lacking the color or another one lacking the left side of the visual field - or with more content - say, one that also includes awareness of the alarm bell that is ringing but I am not hearing it (because I am actually in the middle of a dream). Similarly, my experience flows at a particular speed – each experience encompassing say a hundred milliseconds or so - and there are no superimposed experiences that flow at faster speeds (say, each experience encompassing just one microsecond), or at slower speeds (say, each experience encompassing an entire hour).

**Postulates: Properties that physical mechanisms must have to support consciousness**

To parallel these axioms that capture the essential properties of every experience, IIT proposes a set of postulates concerning the requirements that must be satisfied by physical systems to account for experience (Fig. 3, right).

*Existence*. Experience is generated by a *system of mechanisms in a state*: To *exist* from its own *intrinsic* perspective, independent of external observers, a system must have *cause-effect power* upon itself: its mechanisms must be able to "make a difference" to the probability of its past-future states (Fig. 3, Existence) (Bateson 1972). Cause-effect power is a precondition for something to exist: there is no point in assuming that something exists if nothing can make a difference to it, or if it cannot make a difference to anything (*Alexander's dictum*).[v] To pick up differences and make a difference, physical mechanisms must have two or more internal states, inputs that can influence these states, and outputs that depend on these states. Examples of such mechanisms include neurons and logic gates made of transistors. Moreover, to generate experience, a system of mechanisms must have cause-effect power *within* itself, i.e. *intrinsically*, independent of extrinsic causes and effects.[vi]

*Composition*. The system can be *structured*: elementary mechanisms can be *composed* to specify various "differences that make a difference" to the system (affect the probability of its past-future states). Thus, if a system is composed of elements A, B, and C (Fig. 3, Composition), any subset of elements, including A, B, C; AB, AC, BC; as well as the entire system, ABC, can constitute a mechanism having cause-effect power. Composition thus allows for elementary mechanisms to form distinct higher-order mechanisms (as long as a higher-order mechanism has causes and effects that cannot be reduced to those of its constituting elementary mechanism, see below).

*Information*. A system of mechanisms in a state specifies a *differentiated* conceptual structure: each structure is *the specific way it is* (a *specific* composition of concepts), differing in its *specific* way from other possible ones. A *conceptual structure* is the set of concepts specified by the mechanisms of a system in various compositions. A *concept* is how each mechanism within the system specifies the probability of past-future states of the system (*cause-effect repertoire*). Consider for example, within the system ABC (Fig. 3, Information), the mechanism implemented by element C, an XOR gate with two inputs (A and B) and two outputs (the AND gate B and the OR gate A). If C is OFF, its cause repertoire specifies that, the previous time step, A and B must have been either in the state ON,OFF or in the state OFF,ON, rather than in the other two possible states (ON,ON; OFF,OFF); and its effect repertoire specifies that the next time step B will have to be OFF, rather than ON. Thus, the cause-effect repertoire specifies the cause-effect power of a mechanism in a particular state, and the conceptual structure specifies the cause-effect power of a system of mechanisms. Note that the notion of information in IIT differs substantially from that in communication theory or in common language, but it is faithful to its etymology: information refers to how a system of mechanisms in a state, through its cause-effect power, gives rise to a form ("informs" a conceptual structure) in the space of possibilities.

*Integration*. The conceptual structure specified by the system is *unified*: it is *irreducible* to that specified by non-interdependent sub-systems. Irreducibility can be measured as integrated information (big phi or $\Phi$, a non-negative number), which quantifies to what extent the conceptual structure specified by a system's mechanisms changes if the system is partitioned (cut or reduced) along its minimum partition (the one that makes the least difference). For example, the system in Fig. 3 is integrated, because partitioning it through its weakest link destroys several cause-effect repertoires and changes others (compare the conceptual structure under "information" and under "integration" in Fig. 3). By contrast, if a system of mechanisms can be divided into two sub-systems and the partition makes no difference to the associated conceptual structure, then the whole is reducible to those parts. Being irreducible is another precondition for existence having to do with causation: there is no point in assuming that the whole exists in and of itself, if it has no causal power above and beyond its parts. This postulate also applies to individual mechanisms: a subset of elements can contribute a specific aspect of experience only if its cause-effect repertoire within the system is irreducible by the minimum partition of the mechanism (small phi or $\varphi$).

*Exclusion*. The conceptual structure specified by the system must be *singular*: the one that is *maximally irreducible ($\Phi^{max}$)*. That is, there can be *no superposition* of conceptual structures over elements and spatio-temporal grain. The system of mechanisms that generates a *maximally irreducible conceptual structure* is called a *complex*.[vii] For example, in Fig. 3 (Exclusion), there cannot exist complex ABC, *and* complex AB, AC, *and* BC simultaneously, because complexes cannot overlap. Again, exclusion makes sense with respect to causation, because it avoids multiple causation: if a mechanism specifies a particular cause-effect repertoire within one complex, it cannot *additionally* specify an overlapping cause-effect repertoire as part of other, overlapping complexes, because we would be counting multiple times the difference that mechanism makes. This postulate also applies to individual mechanisms in a complex: a subset of elements in a state can contribute only one aspect to experience - the cause-effect repertoire within the system that is maximally irreducible ($\varphi^{max}$), called a *core concept*. Finally, this postulate also applies to spatio-temporal grain. For example, a mechanism cannot have effects at a fine temporal grain, and additional effects at a coarser grain, otherwise causal exclusion would be violated. On the other hand, if the effects at a coarser grain are more irreducible than those at a finer grain, then the coarser grain of causation excludes the finer one (Hoel, Albantakis et al. 2013).[viii]

### The central identity: Experience as a maximally irreducible conceptual structure

Altogether, the elements of a complex in a state, combined in irreducible mechanisms that specify concepts within the complex, form a *maximally irreducible conceptual structure*, also known as a *quale*. A quale exists in a space called qualia space, whose axes are given by all possible past and future states of the complex. Every concept is a point in the space, which specifies the probability of past and future states of the system, given the state of a particular mechanism within it. The constellation of all concepts together constitutes the "*shape*" of the quale (Fig 4). This leads to the central identity of IIT, which states that a conscious experience is *identical* to a maximally irreducible conceptual structure: the quale completely specifies both its quality (the set of concepts in the quale is the content of consciousness) and its quantity (the value of irreducibility $\Phi^{max}$ of the quale is the level of consciousness). If a system has $\Phi^{max}$ = 0, what it can do as a system is completely reducible to what its parts can do, so it cannot lay claim to existing. If $\Phi^{max}$ is large, the system can do much more than its parts, so it exits in and of itself. More generally, the larger $\Phi^{max}$, the more a system can lay claim to existing, in a fuller sense than lower $\Phi^{max}$ systems. Just *how much* the system exists as such is measured by its $\Phi^{max}$ value, while *which way* it exists is specified by its concepts (irreducible cause-effect repertoires). According to IIT, the quantity and quality of an experience are an *intrinsic* property of a complex of mechanisms in a state – the property of shaping the space of possibilities (past and future states) in a particular way, just as it is intrinsic to a mass to bend space-time around it.[ix]

At any given time, then, consciousness is supported by a set of neuronal mechanisms forming a complex of high $\Phi^{max}$ that specifies a maximally irreducible conceptual structure. The particular set of neurons that form the main complex in our brain may change to some extent from moment to moment, as well as their state – which neurons are firing and which are not. For example, let us assume that while I watch a scene of a movie containing the actress Jennifer Aniston (JA), the main complex in my brain is made up of neurons within certain parts of the cerebral cortex.[x] Every neuron within the complex necessarily shapes the probability of possible past states (causes) and future states (effects) of the complex, depending on how it is connected to the other neurons and on its state (say firing strongly for 100 msec). Thus, a neuron firing strongly in a certain visual area may specify as more likely those past states of the complex that are compatible with the invariant concept "J.A.'s face," as well as certain appropriate future states. Another neuron firing strongly in another visual area may specify that there likely was a horizontal edge in a certain position of the visual field, and so on. Yet other neurons that are part of the complex but are silent may specify that certain past (and future) states are unlikely to have occurred (or to occur), such as those having to do with the invariant concepts "book," "square", and so on. Moreover, combinations of neurons may specify higher-order concepts, such as "J.A. with a red hat sitting on the couch on the left." Note that all the concepts are generated by elements of the complex, specify cause-effect repertoires over elements of the complex, and acquire meaning intrinsically, in relation to the other concepts in the quale, and not by referring to external inputs (J.A. is just as meaningful when daydreaming about her, or in a dream) (Oizumi, Albantakis et al. 2014).

In principle, then, the postulates of IIT offer a way to analyze any system of mechanisms in a particular state and determine whether it constitutes a complex, over which spatial and temporal grain,[xi] and which quale it generates. Furthermore, while in practice it is not possible to determine the quale and $\Phi^{max}$ precisely for a realistic system, it is already possible to employ IIT for prediction, explanation, and extrapolation.

### Predictions

A straightforward experimental prediction of IIT is that the loss and recovery of consciousness should be associated with the breakdown and recovery of the brain's capacity for information integration. This prediction has been confirmed using transcranial magnetic stimulation (TMS) in combination with high-density electroencephalography in conditions characterized by loss of consciousness (Massimini, Ferrarelli et al. 2005, Casali, Gosseries et al. 2013). These include deep sleep, general anesthesia obtained with several different agents, and brain damaged patients (vegetative, minimally conscious, emerging from minimal consciousness, locked-in). If a subject is conscious when the cerebral cortex is probed with a pulse of current induced by the TMS coil from outside the skull, the cortex responds with a complex pattern of reverberating activations and deactivations that is both widespread (integrated) and differentiated in time and space (information rich) (Massimini, Ferrarelli et al. 2005). By contrast, when consciousness fades, the response of the cortex becomes local (loss of integration) or global but stereotypical (loss of information). The *perturbational complexity index* (PCI), a scalar measure of the compressibility of the EEG response to TMS inspired by IIT, decreases distinctly in all the different conditions of loss of consciousness and, critical for a clinically useful device, so far has done so in each individual healthy subject or neurological patient tested (Casali, Gosseries et al. 2013).

Some examples of counterintuitive predictions of IIT are also worth pointing out. One is that a system such as the cerebral cortex may generate experience even if it is nearly silent, a state that is perhaps approximated through certain meditative practices that aim at reaching "pure" awareness without content (Sullivan 1995, Blackmore 2011). This corollary of IIT contrasts with the common assumption that neurons only contribute to consciousness if they are active in such a way that they "signal" or "broadcast" the information they represent and "ignite" fronto-parietal networks (Dehaene and Changeux 2011). That silent neurons can contribute to consciousness is because, in IIT, information is not in the message that is broadcast by an element, but in the shape of the conceptual structure that is specified by a complex. Inactive elements of a complex can affect the probability of possible past and future states as much as active ones can (the dog that did not bark in the famous Sherlock Holmes story). Conversely, if the same neurons were not silent, but pharmacologically or optogenetically turned off (inactivated), they would cease to contribute to consciousness: even though their actual state is the same, they would lack a cause-effect repertoire and thus would not be able to affect the probability of possible past and future states of the complex.

Another prediction is that, if the efficacy of the 200 million callosal fibers through which the two cerebral hemispheres communicate with each other were reduced progressively, there would be a moment at which, for a minimal change in the traffic of neural impulses across the callosum, there would be an all-or-none change in consciousness: experience would go from being a single one to suddenly splitting into two separate experiencing minds (one linguistically dominant), as we know to be the case with split-brain patients (Gazzaniga 2005). This would be the point at which $\Phi^{max}$ for the whole brain would fall below the value of $\Phi^{max}$ for the left and for the right hemisphere taken by themselves.

### Explanations

IIT offers a coherent, principled account for many disparate facts about the NCC. For example, why is consciousness generated by the cerebral cortex (or at least some parts of it), but not by the cerebellum, despite the latter having even more neurons? Why does consciousness fade early in sleep, although the brain remains active? Why is it lost during generalized seizures, when neural activity is intense and synchronous? Why is there no direct contribution to consciousness from neural activity within sensory pathways (the retina) and motor pathways (the motoneurons in the spinal cord), or within neural circuits looping out of the cortex into subcortical structures and back, despite their manifest ability to influence the content of experience?

These and other well-known facts find a parsimonious explanation based on the postulates of IIT. Thus, a prominent feature of the cerebral cortex, responsible for the content of consciousness, is that it is comprised of elements that are functionally specialized and at the same time can interact rapidly and effectively. This is the kind of organization that yields a comparatively high value of $\Phi^{max}$. Instead, the cerebellum is composed of small sheet-like modules that process inputs and produce outputs largely independent of each other (Cohen 1998, Bower 2002). Simulations also show that input and output pathways, while capable of affecting the main complex and being affected by it, can remain excluded from it, because they are not part of a local maximum of integrated information. The same applies to loops that may exit the main complex and reenter it. Other simulations show that $\Phi^{max}$ is low when the effective connectivity among a set of elements is weak or is organized in homogeneous manner. Indeed, as was mentioned above, when consciousness fades during deep slow wave sleep or in certain states of general anesthesia, the interactions among different cortical regions become weaker and highly stereotypical, as they do during generalized epileptic seizures.

### Extrapolations

Finally, the more the postulates of IIT are validated in situations in which we are reasonably confident about if and how consciousness changes, the more we can use the theory to extrapolate and make inferences about situations where we are less confident – brain-damaged patients, newborn babies, alien animals, complicated machines, and other far-fetched scenarios, as we shall consider next.

### Everywhere?

In the "Canticle of the Creatures," Saint Francis addressed animals, flowers, and even stones as if endowed with soul (psyche), and praised God for all His creatures: mother earth, brother sun, sister moon, the stars, the air, water, and fire. And he was not alone. Some of the brightest minds in the West embraced some form of the ancient philosophical doctrine of panpsychism, starting with the Presocratics and Plato. The Renaissance philosophers Patrizi, Bruno, Telesio, and Campanella took the position that matter and soul are one substance. Later, Spinoza, Leibniz, Schopenhauer and, closer to modern times, James, Whitehead, Russell, and Teilhard de Chardin espoused panpsychist notions (Skrbina 2009, Chalmers 2013). Strawson (2006) (Strawson and Freeman 2006) is the best-known contemporary defender of panpsychism. Eastern traditions, such as Buddhism, have always emphasized the continuity of consciousness across creatures.

Materialism, or its modern offspring, physicalism, has profited immensely from Galileo's pragmatic stance of removing subjectivity (mind) from nature in order to describe and understand it objectively. But it has done so at the cost of failing to deal with the central aspect of reality – experience itself. Unlike idealism, which does away with the physical world, or dualism, which accepts both in an uneasy marriage, panpsychism is elegantly unitary: there is only one substance, all the way up from the smallest entities to human consciousness and maybe to the World Soul (*anima mundi*). But panpsychism's beauty has been singularly barren. Besides claiming that matter and mind are one thing, it has little constructive to say and offers no positive laws explaining how the mind is organized and works.

IIT was not developed with panpsychism in mind (*sic*). However, in line with the central intuitions of panpsychism, IIT treats consciousness as an intrinsic, fundamental property of reality. IIT also implies that consciousness is graded, that it is likely widespread among animals, and that it can be found in small amounts even in certain simple systems. Unlike panpsychism, however, IIT clearly implies that consciousness is not ubiquitous. Moreover, IIT offers a solution to several of the conceptual obstacles that panpsychists never properly resolved, like the problem of aggregates (or combination problem, (James 1890, Chalmers 2013)). It also explains why consciousness is adaptive, and can account for its quality.

### Consciousness is an intrinsic property

The axioms and postulates of IIT say that consciousness is an intrinsic, observer-independent property of certain mechanisms in a state - how they shape the space of possibilities in their past and their future. An analogy is mass, which can be defined by how it curves space-time around it. Like mass, experience cannot be reduced to more elementary properties. And like some elementary particles have mass or charge and others do not, and if they have it, they have it in a particular way (positive or negative charge), so for experience: some entities have it, while others do not, and if they have it, they have it in a certain way. Except that in the case of experience the entities having the property are not elementary particles, but complexes of mechanisms, and experience comes not in two, but in a trillion varieties. Thus, for IIT, we happen to find ourselves in a universe in which experience is one of the elementary properties of certain causal systems. Asking why this should be so or whether we can imagine a universe in which this is not true is akin to asking why our universe obeys the laws of quantum mechanics or whether we can image a universe in which quantum mechanics does not hold. In this general sense, at least, IIT is not at odds with panpsychism.

### Consciousness comes in various qualities

Unfortunately, panpsychism is mute when it comes to explaining the way any one conscious experience feels - why the perception of red feels different from that of blue and why colors are experienced as different from tones. Instead, at least in principle, IIT says exactly what determines the quality of an experience – what makes it the particular way it is: an experience is identical to the maximally irreducible conceptual structure (the quale) generated by a main complex, in our case one made up by a set of neurons in a particular state. This is like a shape, a constellation in a fantastically high-dimensional qualia space, which specifies how the neurons of the main complex, in various combinations, give form to the space of possible past and future states of the complex (Fig. 4). Different experiences – every different scene in a movie or in a dream - correspond to different shapes. Except that this shape, unlike the shapes of objects, is the shape within, the shape of the experience itself. It is the voice in the head, the light inside the skull. It is everything I will ever know of the world. It is, ultimately, my only reality.

### Consciousness is adaptive

IIT takes no position on the function of experience as such - similar to physics not having anything to say about the function of mass or charge. However, by identifying consciousness with integrated information, IIT can account for why it evolved, another aspect about which panpsychism has nothing to say. In general, a brain having a high capacity for information integration will better match an environment with a complex causal structure varying across multiple time scales, than a network made of many modules that are informationally encapsulated. Indeed, artificial life simulations ("animats") of simple Braitenberg-like vehicles that have to traverse mazes and whose brains evolve, over 60,000 generations, by natural selection, show a monotonic relationship between (simulated) integrated information and adaptation (Edlund, Chaumont et al. 2011, Joshi, Tononi et al. 2013). That is, the more adapted individual animats are to their environment, the higher the integrated information of the main complex in their brain. Thus, evolution by natural selection gives rise to organisms with high $\Phi^{max}$ because they are more adept at exploiting regularities in the environment than their less integrated competitors.

### Consciousness is graded

IIT does side with the panpsychist intuition that consciousness may be present across the animal kingdom, and even beyond, but in varying degrees. Everything else being equal, integrated information, and with it the richness of experience, is likely to increase as the number of neurons and the abundance of their interconnections grow, although sheer number of neurons is not a guarantee, as shown by the cerebellum. It is also likely that consciousness is graded across the lifetime of any one organism. In us it becomes richer as we grow from a baby to an adult whose brain has fully matured and becomes more functionally specialized. It can also wax and wane when we are highly alert or drowsy, intoxicated by drugs or alcohol, or become demented in old age. This is illustrated schematically in Fig. 5A, where a set of "cortical" areas are integrated into a main complex of "high" $\Phi^{max}$ when the inter-areal connections are strong, undergo a reduction in $\Phi^{max}$ when connection strength is reduced by neuromodulatory changes (simulated as an increase in noise), and finally breaks down into small complexes of low $\Phi^{max}$.

A counterintuitive corollary of IIT is that even circuits as simple as a single photodiode hooked up to a 1-bit memory can have a modicum of experience (Oizumi, Albantakis et al. 2014) (see also Fig. 5A, right panel). It feels like something to be this circuit; it is conscious of one thing – the distinction between this and not this (not of light or dark, because that requires many more concepts). This strongly violates our intuitions about consciousness. But consider that normal matter at -272.15° C, one degree above absolute zero, still contains some heat that, in theory, could be extracted. However, in practice its temperature is as cold as it gets. Similarly, there may well be a practical threshold for $\Phi^{max}$ below which people do not report feeling much of anything, but this does not mean that consciousness has reached its absolute zero. Indeed, when we fall into a deep, dreamless sleep and don't report any experience upon being awoken, our sleeping brain is still not fully disconnected and some complex within it will likely have a $\Phi^{max}$ value greater than zero, yet that may not amount to much compared to that of our rich, everyday experience.

### Aggregates are not conscious

"Take a sentence of a dozen words, and take twelve men and tell to each one word. Then stand the men in a row or jam them in a bunch, and let each think of his word as intently as he will; nowhere will there be a consciousness of the whole sentence." This is how William James illustrated the combination problem of panpsychism (James 1890). Or take John Searle: "Consciousness cannot spread over the universe like a thin veneer of jam; there has to be a point where my consciousness ends and yours begins" (Searle 2013). Indeed, if consciousness is everywhere, why should it not animate the United States of America? IIT deals squarely with this problem by stating that only maxima of integrated information count. Consider two people talking: within each brain, there will be a main complex – a set of neurons that form a maximally irreducible whole with definite borders and a high value of $\Phi^{max}$. Now let the two speak together. They will now form a system that is also irreducible ($\Phi$ is greater than zero) due to their interactions. However, it is not maximally irreducible, since its value of integrated information will be much less than that of each of the two main complexes it contains. According to IIT, there should indeed be two separate experiences, but no superordinate conscious entity that is the union of the two. In other words, there is nothing it-is-like-to-be two people, let alone the 300 plus million citizens making up the USA.[xii] Again, this point can be exemplified schematically by the system of Fig. 5A, right panel. While the five small complexes do interact, forming a larger integrated system, the larger system is not a complex: by the exclusion postulate, only the five smaller complexes exists, since they are local maxima of integrated information ($\Phi^{max}$ = 0.19), while the larger system is not a complex ($\Phi$ = 0.03). Worse, a dumb thing with no distinguishable states, say a grain of sand for the sake of the argument, has no experience whatsoever. Heaping a large number of such zero-$\Phi$ systems on top of each other would not increase their $\Phi$ to a non-zero value; it does not feel like anything to be a sand dune. Aggregates have no consciousness.

### Complicated systems can be unconscious

A second class of zero-$\Phi$ systems are purely feed-forward computational networks in which one layer feeds the next one without any recurrent connections. According to IIT, a feed-forward network does not have a cause repertoire within the system itself, since its input is imposed from outside the system, nor does it have an effect repertoire, since its output does not feed back to any element within the network. Yet feed-forward networks, such as the those used in deep learning, perform plenty of useful computational functions, such as finding faces or cats in images (Le, Ranzato et al. 2011), labeling images, reading zip codes and detecting credit card fraud. From the point of IIT, such networks are zombies, carrying out tasks unconsciously (Koch and Crick 2001).

This has a rather startling consequence. Consider that any neural network with feedback circuits can be mapped onto a purely feed-forward network in such a manner that the latter approximates its input-output relationships (for computations bounded by a maximal time-step; (Hornik, Stinchcombe et al. 1989). That is, for the same inputs, the two networks will yield the same output (in general, the equivalent feed-forward network will have many more nodes and connection than the feedback network). Therefore, a purely feed-forward system that were able to replicate the input-output behavior of the human brain (under the limited time-step constraint), while behaviorally indistinguishable from us, and certainly capable of passing the Turing test, would have zero $\Phi$, and would thus be a "perfect" zombie. A simple example of two functionally equivalent systems, one with recurrent connections and non-zero $\Phi$, and one purely feed-forward with zero $\Phi$, is show in Fig. 5B (Oizumi, Albantakis et al. 2014).

In people and organisms that evolved through natural selection, their input-output behavior, as assessed by the BCC, offers a good first guess about the presence of consciousness. As demonstrated by the example in Fig. 5B, this may not always be the case for radically different computational architectures. In the general case, and certainly with machines, it becomes absolutely essential to consider the internal circuitry – not just what the machine does, but how it does so. This also means that there cannot be an ultimate Turing Test for consciousness (although, there may be some practical CAPTCHA-like tests; (Koch and Tononi 2011)). According to many functionalist notions (Dennett 1993), if a machine reproduces our input-output behavior in every circumstance, it would have to be granted consciousness just as much as us. IIT could not disagree more.

### Simulations of conscious neural systems can be unconscious

Finally, what about a computer whose software simulates in detail not just our behavior, but even the biophysics of synapses, axons, neurons and so on, of the relevant portion of the human brain (Markram 2006)? Could such a digital simulacrum ever be conscious? Functionalism again would say yes, even more forcefully. For in this case *all* the relevant interactions within our brain, not just our input-output behavior, would have been replicated faithfully. Why should we not grant to this simulacrum the same consciousness we grant to a fellow human? According to IIT, however, this would not be justified, for the simple reason that the brain is real, but a simulation of a brain is virtual (including, *sensu stricto,* its simulated value of $\Phi^{max}$). For IIT, consciousness is an intrinsic property of certain systems of mechanisms, one that requires having *real* causal power, specifically the power of shaping the space of possible past and future states in a maximally irreducible way. In the same way, mass is an intrinsic property of systems of particles, a property that has real causal power, specifically that of bending space-time. Therefore, just like a computer simulation of a giant star will not bend space-time around the machine, a simulation of our conscious brain will not have consciousness.[xiii] Of course, the physical computer that is running the simulation is just as real as the brain. However, according to the principles of IIT, one should analyze its real physical components - identify elements, say transistors, define their cause-effect repertoires, find concepts, complexes, and determine the spatio-temporal scale at which $\Phi$ reaches a maximum. In that case, we suspect that the computer would likely *not* form a large complex of high $\Phi^{max}$, but break down into many mini-complexes of low $\Phi^{max}$ (due to the small fan-in and fan-out of digital circuitry, Fig. 5C), existing at the very fast temporal scale of the computer clock.[xiv]

### Conclusion

In summary, there are some aspects of IIT that definitely do not fit with panpsychism, and others that vindicate some of its intuitions. In this respect, it is natural to consider how one should regard some of the inferences derived from IIT for which it is hard even to imagine a direct test at the present time. Our position is that, as is often the case in science,[xv] a theory is first tested and validated in situations that are close to ideal, and then extrapolated to more remote cases. Ideally, whether consciousness varies with integrated information, and other predictions of IIT, would first be validated *here* – on my own consciousness: for example, does $\Phi^{max}$, collapse when I undergo general anesthesia or a seizure, or when I fall into dreamless sleep, and return to high values when I dream? Then one can extrapolate to *there*, at first in situations involving other healthy humans, then in slightly more difficult cases, say monkeys with a brain similar to ours who are trained to give reports similar to ours. Finally, in so far as the theory has been validated and has shown good predictive and explanatory power, one can try and extrapolate to *everywhere*, unresponsive patients with just a small "island" of functioning brain tissue, newborn babies, animals very different from us, photodiodes, machines, and computer simulations. After all, often in science the best we can do is to draw inferences about unknown instances based on a theory that works well in many known instances. And that is much better than to make arbitrary claims or to draw no inference whatsoever.

### Acknowledgements

We thank Larissa Albantakis, Melanie Boly, Chiara Cirelli, Lice Ghilardi, and Marcello Massimini, for their many contributions to the work presented here.

**Table 1. Some terms used in Integrated Information Theory (IIT)**

**Axioms**: Self-evident truths about consciousness. The only truths that, with Descartes, cannot be doubted and do not need proof. In IIT 3.0, there are five such axioms - existence, composition, information, integration, and exclusion (Fig. 3).

**Postulates**: Assumptions, derived from axioms, about the physical substrates of consciousness (mechanisms must have cause-effect power, be irreducible, etc.), which can be formalized and form the basis of the mathematical framework of IIT. There are 5 postulates, matching the five axioms (Fig. 3).

**Element**: An elementary component of a system, for example a neuron in the brain, or a logic gate in a computer.

**Mechanism**: Any subset of elements within a system, including the system itself, which has cause-effect power within the system.

**Cause-effect repertoire**: The probability distribution of potential past and future states of a system as informed by a mechanism in its current state.

**Integrated information ($\varphi$)**: Information that is generated by a mechanism above and beyond the information generated by its (minimal) parts. $\varphi$ measures the integration or irreducibility of the cause-effect repertoire specified by a mechanism.

**MIP** (minimum information partition): The partition that makes the least difference (in other words, the minimum "difference" partition).

**Complex**: A set of elements within a system that generates a local maximum of integrated conceptual information $\Phi^{max}$. Only a complex exists as an entity from its own intrinsic perspective.

**Concept**: A mechanism and the maximally irreducible cause-effect repertoire it specifies, with its associated value of integrated information $\varphi^{max}$. The concept expresses the cause-effect power of a mechanism within a complex.

**Conceptual structure**: The set of all concepts specified by a system set with their respective $\varphi^{max}$ values, which can be plotted as a constellation of concepts in concept space.

**Concept space**: Concept space is a high dimensional space with one axis for each possible past and future state of the system in which a conceptual structure can be represented.

**Integrated conceptual information ($\Phi$)**: Conceptual information that is generated by a system above and beyond the conceptual information generated by its (minimal) parts. $\Phi$ measures the integration or irreducibility of a constellation of concepts (integration at the system level), a non-negative number.

**Quale**: The maximally integrated conceptual structure generated by a complex in a state (synonymous with constellation in qualia space).

**Qualia space**: If a set of elements forms a complex, its concept space is called qualia space.

**Figures**

**Figure 1**: Behavioral (BCC) and neuronal correlates of consciousness (NCC). The top row shows a schematic diagram of a binocular rivalry experiment. A horizontal red grating is shown to the left eye and a vertical green grating to the right eye throughout the experiment (courtesy of Naotsugu Tsuchiya and Olivia Carter). The subject does not see a juxtaposition of both stimuli but experiences either the red grating or the green one, switching back and forth every few seconds. Even if the stimuli do not change, what one sees consciously does, as is inferred by the subject's report. The bottom row shows the results of such an experiment using magnetoencephalography (MEG), in which the red grating was flashed at one frequency and the green one at another. Yellow indicates areas of the cortex (seen from the top) that had more power at the frequency of the red grating when it was experienced than when it was not. The cyan lines indicate increased coherence (synchronization) between distant brain regions associated with experiencing the grating (from Tononi et al., 1998).

**Figure 2**: Six instances in which it becomes progressively more difficult to infer the existence of consciousness, since the behavioral repertoire and the underlying mechanisms (brains) differ substantially from that of typical persons able to speak about their experiences (Fig. 1).

**Figure 3**: Axioms and Postulates of Integrated Information Theory (IIT). The illustration is a colorized version of Ernst Mach's "View from the left eye" (Mach 1959). See also the mechanism in Fig. 4.

**Figure 4**: A didactic example of how to calculate the quality and quantity of consciousness given a mechanism in a state. On the upper left are three gates with binary states (either ON or OFF: ABC = 100; see also Fig. 3) that are wired together as shown. An analysis based on the postulates of IIT (Oizumi, Albantakis et al. 2014) reveals that the system forms an irreducible complex. The complex in its present state specifies a quale - a maximally irreducible conceptual structure. The quale is presented both as the set of maximally irreducible cause-effect repertoires (concepts) specified by each mechanism (top) and as a 2-D projection in which each concept is a "star" in concept space (bottom). Concept space is a high-dimensional (here, 2x8 dimensions) space in which each axis is a possible past (in blue) and future (in green) state of the complex, and the position along the axis is the probability of that state. Each concept is a star whose position indicates how it affects the probability of past and future states of the system (its cause-effect repertoire, which specifies what the concept contributes to experience) and its size ($\varphi^{max}$) measures how irreducible the concept is (how much it contributes to experience). In IIT, $\Phi^{max}$ - a non-negative number - measures the irreducibility of the entire quale, how much consciousness there is – the quantity of experience. The "shape" of the quale (constellation of stars) is identical to the quality of the experience. Different shapes correspond to different experiences: they feel the way they do - red feeling different from blue or from a headache - because of the distinct shapes of their qualia.

**Figure 5**: IIT makes several predictions about which system can experience anything (how much and in which way) and which systems, even complicated ones, have no experience, are "in the dark". IIT implies that consciousness is graded (A); that aggregates are not conscious (A, right panel); that strictly feed-forward systems are not conscious (B, right panel), even if they are functionally equivalent in terms of their input-output operations to feedback networks that are conscious (B, left panel); that even accurate biophysical simulations of the human brain running on digital machines would not be conscious like us, but constitute mere aggregates of much simpler systems (transistors and the like) having minimal $\Phi^{max}$ (C). The last row (C) shows, from left to right, a human brain (Allen Institute), the IBM Blue Gene P supercomputer, a columnar model of mouse cortex (Blue Brain Project), and a scanning electron micrographic cross-section of 4 NMOS INTEL transistors in a grid.

**Notes**

[i] Note that we consider self-consciousness, highly developed in adult humans, to be a subclass of conscious experiences. Likewise, the feeling of freely willing an action – such as raising one's arm - sometimes also referred to as *agency* (Fried, Katz et al. 1991, Wegner 2002) – is another subclass of conscious experiences. While their content differs from the content associated with feeling pain or seeing red, subjectivity is common to all.

[ii] Consciousness can be dissociated from many other cognitive processes that have traditionally been closely linked to it, including memory, emotions and selective attention (for reviews see (Tononi and Koch 2008, Koch and Tsuchiya 2012). It can persist if the recall of long-term memories is impaired, it can be present in patients who lack affect, and it can be dissociated from attention. The last point is particularly counterintuitive but is well-supported – subjects can attend to invisible objects (Tsuchiya and Koch 2014). The extent to which it is possible to become conscious of something without also attending to it is more controversial (Cohen, Cavanagh et al. 2012, Tsuchiya and Koch 2014).

[iii] Not to mention the question of whether it feels-like-something to be a Venus flytrap or a single-cell organism.

[iv] If it is not outright wrong, IIT most likely will have to be refined, expanded, and adjusted. However, in its current form (IIT 3.0), it explains and predicts a wide range of phenomena, including a number of counterintuitive predictions amenable to empirical falsification. For the latest formulation of the theory, see (Oizumi, Albantakis et al. 2014); for earlier versions, see (Tononi 2004, Balduzzi and Tononi 2008, Tononi 2008, Balduzzi and Tononi 2009, Tononi 2012); for a literary account, see (Tononi 2012). The main differences between IIT 3.0 and earlier versions are listed in (Oizumi, Albantakis et al. 2014).

[v] For example, the notion of the aether was introduced in the late 19$^{th}$ century to explain the propagation of light. When more and more experiments concluded that, whatever the aether might be, it had no effects whatsoever, it finally fell under Occam's razor, and it plays no role in modern physics.

[vi] However, mechanisms outside a system can serve as fixed boundary conditions (Oizumi, Albantakis et al. 2014).

[vii] Importantly, this may be a macro- rather than a micro-spatio-temporal scale (Hoel, Albantakis et al. 2013). For example, the relevant level for human consciousness is likely to be neurons at the scale of 100 millisecond rather than molecules at the nanosecond scale.

[viii] Requiring that only the maximum of Φ over elements, spatial, and temporal grain must be considered is not exceptional in science: many of the laws of physics are formulated as extremum principles, e.g. the principle of least action.

[ix] IIT postulates that experience is a fundamental, intrinsic property of organized matter, yet supervenient upon the physical. While different physical states can give rise to the same conscious experience (*metamers*) (Palmer 1999), different experiences must be caused by differences in the underlying physical mechanism. Note that IIT is compatible with quantum mechanics. In principle, Φ and related quantities can be assessed also in quantum system, although it has been suggested that at the quantum level Φ values may be very small (Tegmark 2014).

[x] Here we do not elaborate about particular cortical areas, cortical layers, or particular population of neurons.

[xi] The exclusion postulate, by requiring that the set of mechanisms that generate one particular experience do so over the time window at which Φ reaches a maximum, would seem to imply that, to avoid multiple causation, the next experience should be generated over a non-overlapping time window, as long as an overlapping set of mechanisms is involved. The seemingly continuous "stream" of consciousness would actually be constituted by a discrete succession of "snapshots," in line with some psychophysical evidence (Stroud 1956)'(Crick and Koch 2003, VanRullen and Koch 2003, VanRullen 2014). Note that each snapshot has motion and other dynamic percepts associated with it.

[xii] By the same token, the exclusion postulate predicts a scenario that is the mirror image of the prediction that consciousness will suddenly split in two when the corpus callosum is "cooled" below a critical point: if two people speaking were to increase their effective causal interactions by some, yet to be invented, direct brain-to-brain connectivity booster, to the point where the $\Phi^{max}$ of the two interacting brains would exceed $\Phi^{max}$ of the individual brains, their individual conscious mind would disappear and its place would be taken by a new über-mind that subsumes both.

[xiii] A similar point was made by John Searle with his Chinese Room Argument (Searle 1980) and by Leibniz 300 years earlier with his mill (Leibniz, Montgomery et al. 2005).

[xiv] In the extreme case, any digital computer running software can ultimately be mimicked by a Turing Machine with a large state-transition matrix, a moving head that writes and erases, and a very, very long memory tape – in that case, causal power would reside in the moving head that follows one out of a few instructions at a time. On the other hand, there is no reason why a hardware-level, neuromorphic model of the human brain system that does not rely on software running on a digital computer, could not approximate, one day, our level of consciousness (Schmuker, Pfeil et al. 2014).

A related question has to do with the Internet and whether it could be conscious (Koch 2014). One way to think about this is to assume that each computer connected to the Internet is an element having real causal power at a macro-level. For example, each computer could send an ON-line signal when it is ON and an OFF-line signal when it is OFF. One could then make sure that each computer increased or decreased the likelihood of being ON depending on how many ON-signals it received. In principle, this kind of organization could be arranged so that it gives rise to a complex of high Φ, although this is certainly not the way the Internet works right now. On the other hand, if one considers the micro-elements inside each computer (say its transistors) as having real causal power, we are back to the situation in which they most likely would not form any large complex within each computer, let alone across connected computers.

[xv] A well-known instance of such an extrapolation is the inference of singularities in space-time due to the extreme mass of a stellar object. Such black holes were pure conjectures, based on a solution of Einstein's theory of General Relativity, until there were subsequently confirmed observationally.

Fig. 1

**BCC**

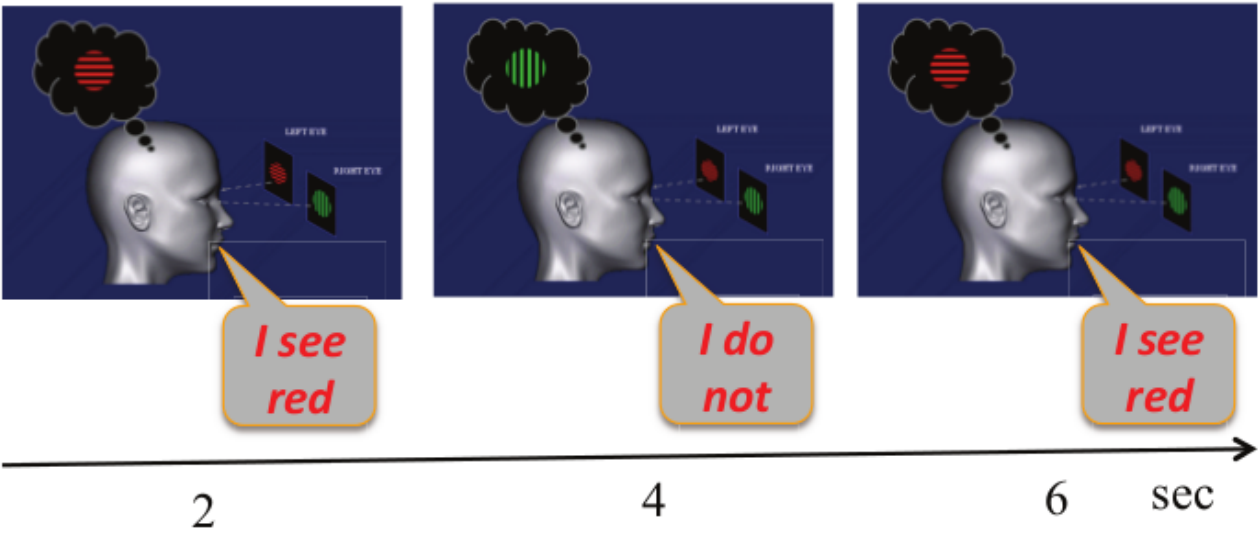

**NCC**

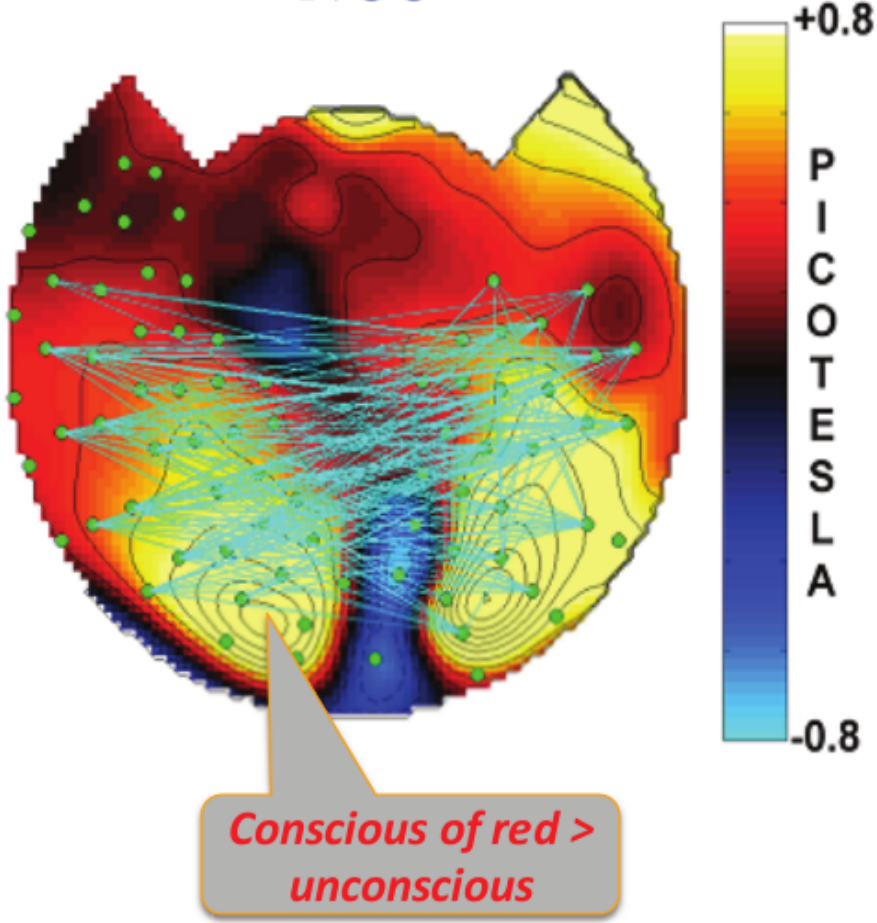

Fig. 2

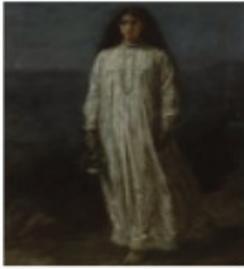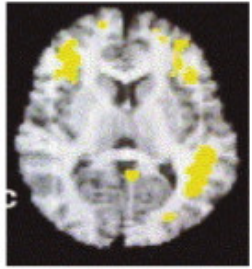
Sleepwalking

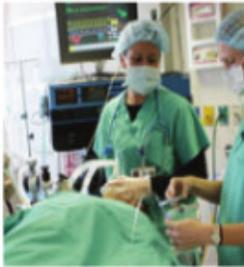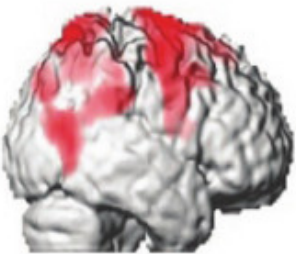
Ketamine anesthesia

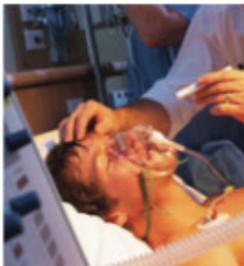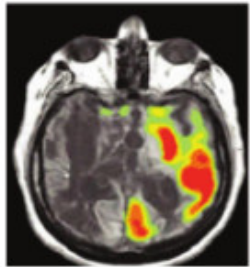
Brain "islands," vegetative patient

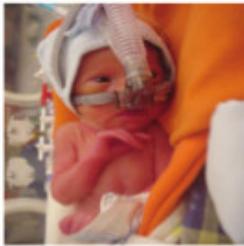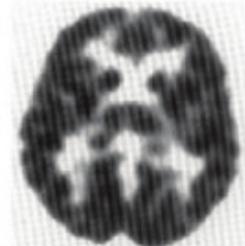
Pre-term infant

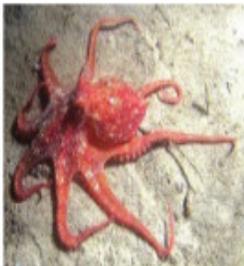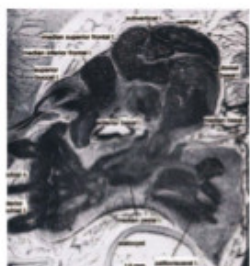
Octopus

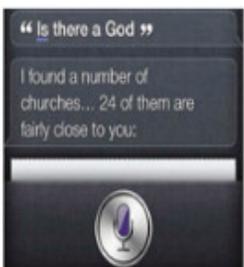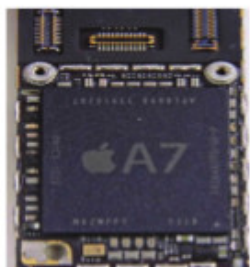
Apple Siri

## Axioms | Postulates

### Existence

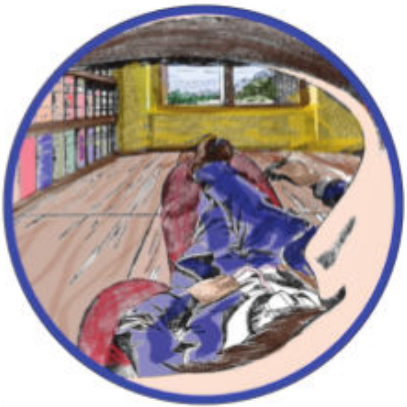

Consciousness **exists**: each experience exists from its own *intrinsic* perspective, independent of external observers

Experience is generated by a system of mechanisms in a state: to **exist** from its *intrinsic* perspective, independent of external observers, the system must have cause-effect power within itself (its mechanisms must "make a difference" to the probability of its past and future states)

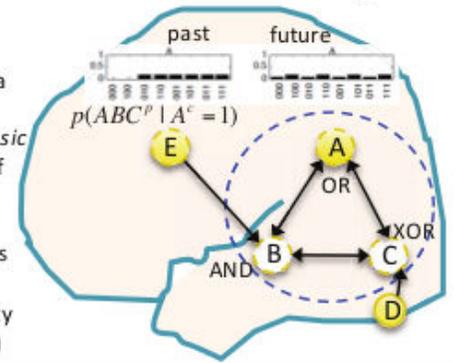

### Composition

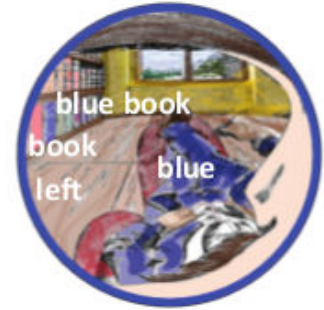

Consciousness is **structured**: each experience is composed of many phenomenological distinctions

The system can be **structured**: elementary mechanisms can be *combined* into higher-order mechanisms to specify various "differences that make a difference" to the system

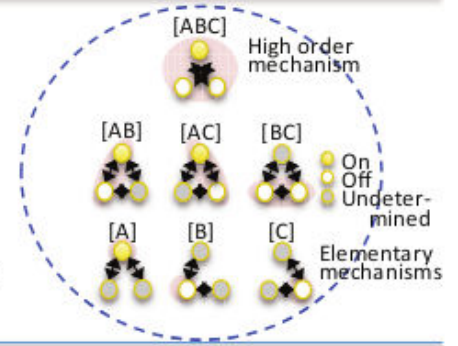

### Information

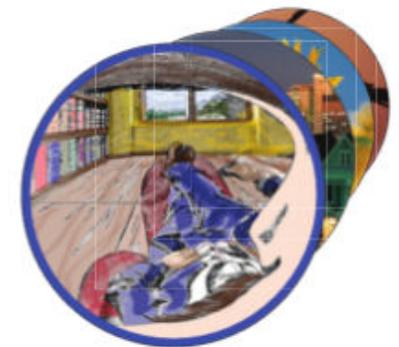

Consciousness is **differentiated**: each experience is *the specific way it is* (a specific composition of phenomenological distinctions), differing in its specific way from other possible experiences

The system of mechanisms in a state must specify a **differentiated** conceptual structure. A *concept* is how each mechanism specifies the probability of past-future states of the system (*cause-effect repertoire*). A *conceptual structure* is the set of all concepts specified by a system's mechanisms, and it expresses how the system's mechanisms give "form" to the space of possibilities. Each conceptual structure is *the specific way it is* (a specific composition of concepts), differing in its specific way from other possible structures

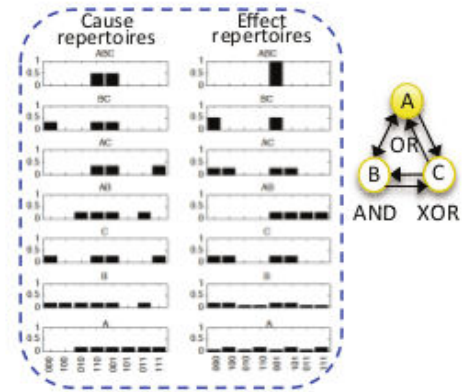

### Integration

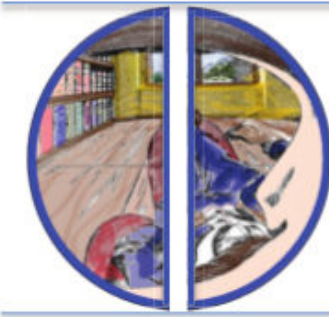

Consciousness is **unified**: each experience is *irreducible* to non-interdependent components

The conceptual structure specified by the system must be **unified**: it must be *irreducible* to that specified by non-interdependent sub-systems ($\Phi > 0$) (across its weakest link MIP = minimum information partition)

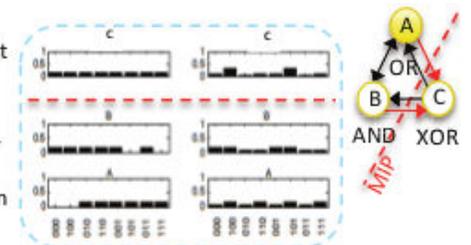

### Exclusion

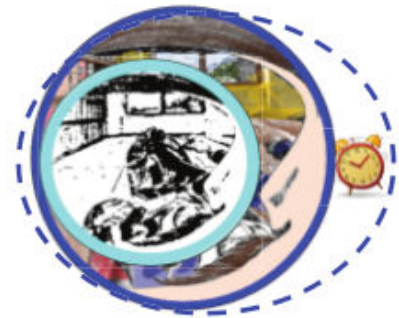

Consciousness is **singular**, in content and spatio-temporal grain (there is *no superposition* of multiple experiences, with less or more content, flowing at once at faster or slower speeds)

The conceptual structure specified by the system must be **singular**: the one that is maximally irreducible ($\Phi^{max}$): there can be *no superposition* of conceptual structures over elements and spatio-temporal grain

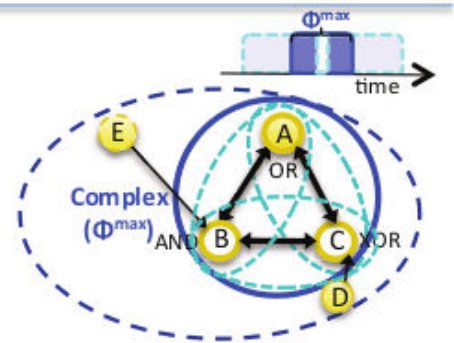

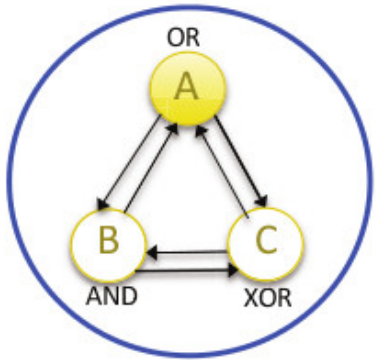

A complex

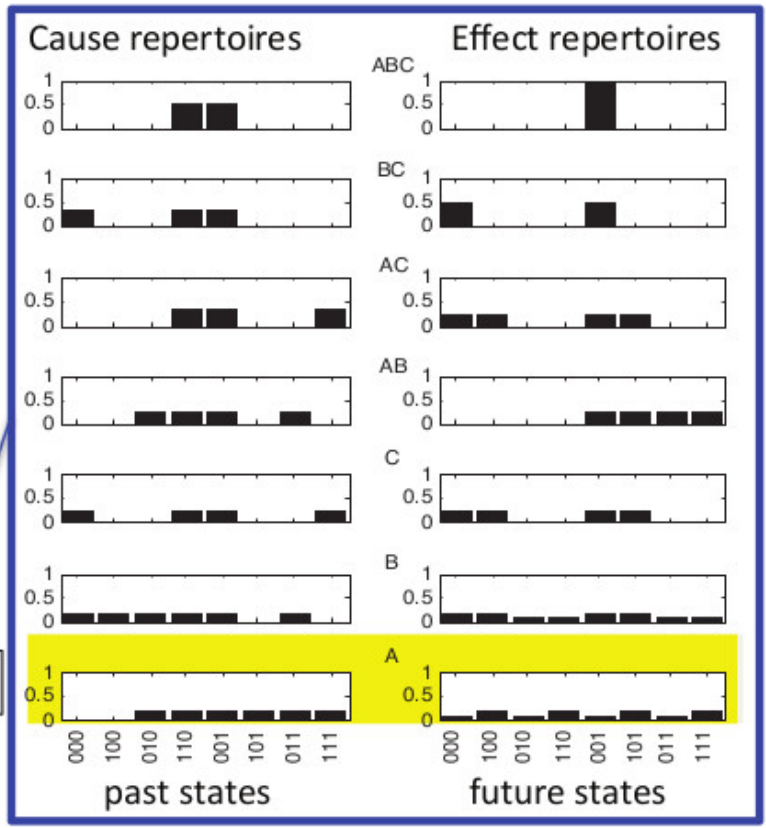

Cause repertoires | Effect repertoires

concept

A quale: A maximally irreducible conceptual structure in concept space

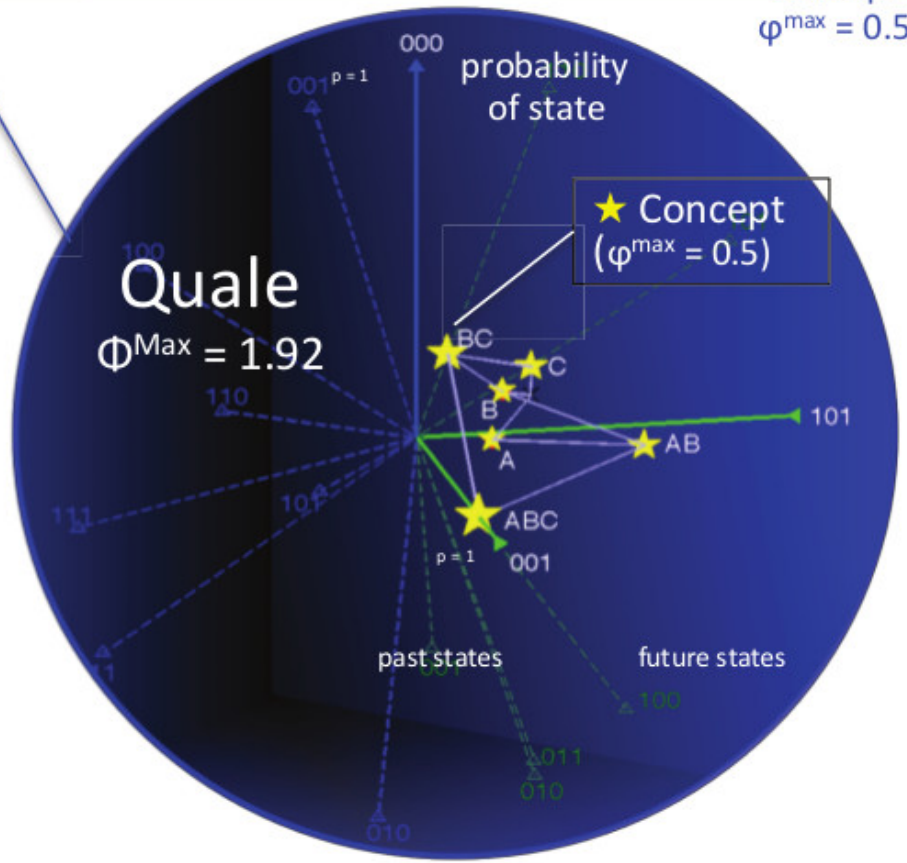

Quantity:
irreducibility ($\Phi^{max}$)
of the quale

Quale
$\Phi^{Max} = 1.92$

Concept
($\varphi^{max} = 0.5$)

Quality:
"shape" of the quale

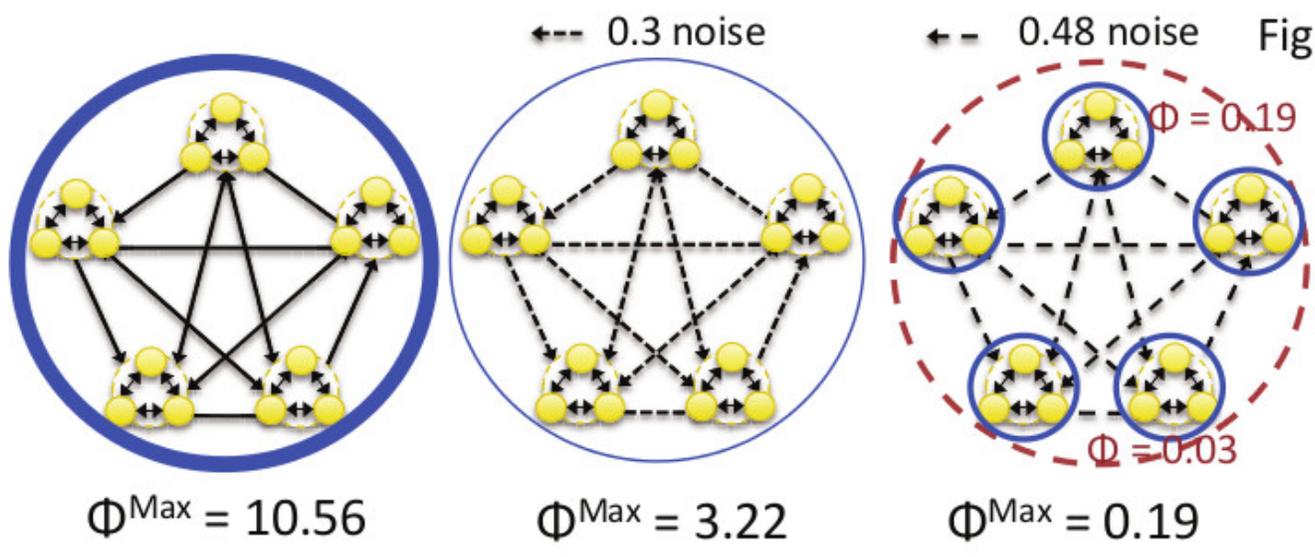
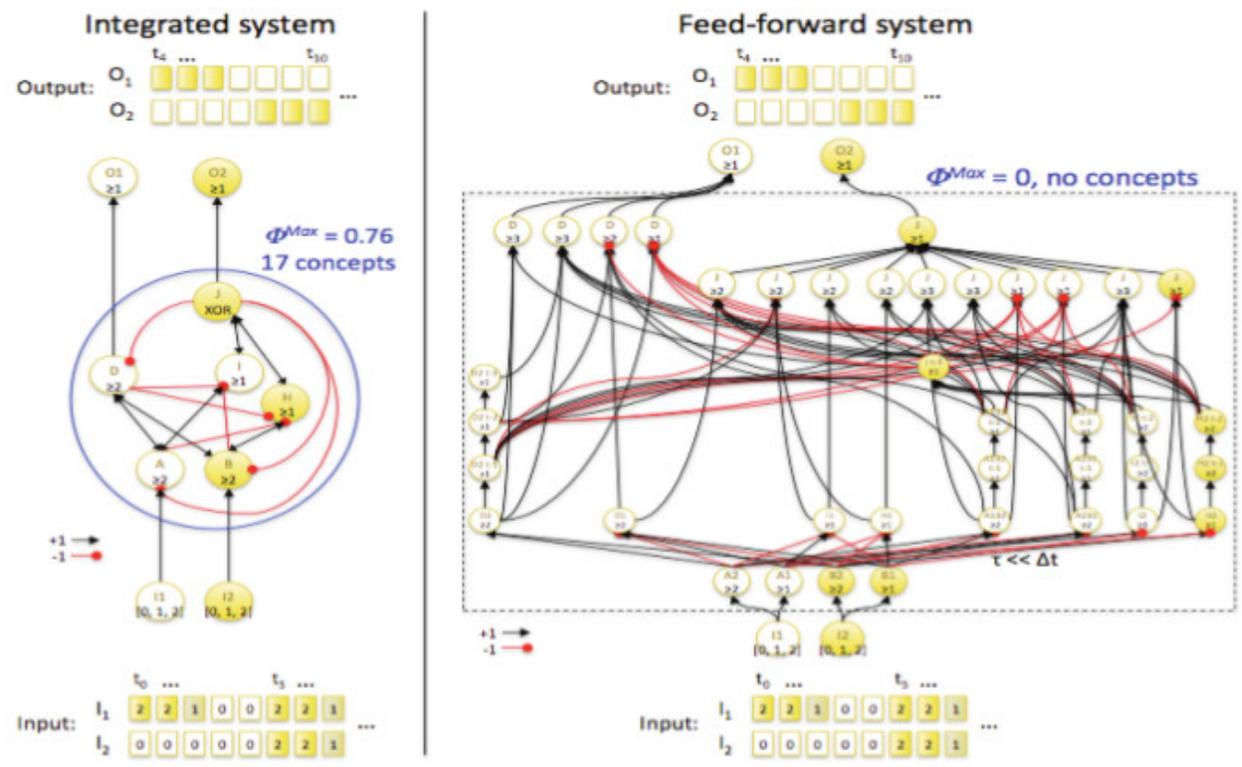
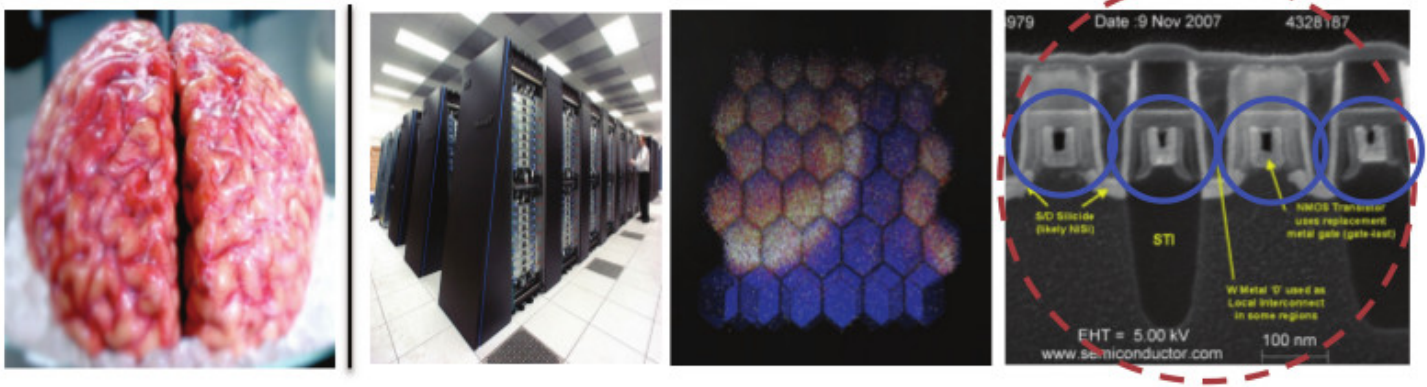

Fig. 5